# 5G QoS: Impact of Security Functions on Latency

Sebastian Gallenmüller*, Johannes Naab*, Iris Adam†, Georg Carle*
*Technical University of Munich, †Nokia Bell Labs
*{gallenmu, naab, carle}@net.in.tum.de, †iris.adam@nokia-bell-labs.com

*Abstract*—Network slicing is considered a key enabler to 5th Generation (5G) communication networks. Mobile network operators may deploy network slices—complete logical networks customized for specific services expecting a certain Quality of Service (QoS). New business models like Network Slice-as-a-Service offerings to customers from vertical industries require negotiated Service Level Agreement (SLA) contracts, and network providers need automated enforcement mechanisms to assure QoS during instantiation and operation of slices. In this paper, we focus on ultra-reliable low-latency communication (URLLC). We propose a software architecture for security functions based on off-the-shelf hardware and open-source software and demonstrate, through a series of measurements, that the strict requirements of URLLC services can be achieved. As a real-world example, we perform our experiments using the intrusion prevention system (IPS) Snort to demonstrate the impact of security functions on latency. Our findings lead to the creation of a model predicting the system load that still meets the URLLC latency requirement. We fully disclose the artifacts presented in this paper including pcap traces, measurement tools, and plotting scripts at https://gallenmu.github.io/low-latency.

## I. INTRODUCTION

The flexibility and adaptability of 5G are considered its main features enabling the creation of dedicated wireless networks customized for specific applications with a certain level of QoS. The International Telecommunication Union (ITU) identified three distinct services for 5G networks [1]: *enhanced mobile broadband* (eMMB): a service comparable to LTE networks optimized for high throughput; the *massive machine type communication* (mMTC): a service designed for spanning large IoT networks optimized for a large number of devices with low power consumption; and the *ultra-reliable low-latency communication* (URLLC): a service for safety-critical applications requiring high reliability and low latency. These different services can be realized by *slicing* the network into distinct independent logical networks which can be offered as a service adhering to customer-specific SLAs, called Network Slice-as-a-Service. A cost-efficient way to realize network slices is the shared use of network resources among customers, e.g. virtualization techniques used on off-the-shelf servers. This makes virtualization and its implications on performance one of the crucial techniques used for 5G. Virtualization is the natural enemy of predictability and low latency [2] posing a major obstacle when realizing URLLC. In this paper, we investigate if and how the seemingly contradictory optimization goals virtualization and resource sharing on the one side and low latency and high predictability on the other side can go together. The goals of our investigation are threefold: *(i)* creating a low-latency packet processing architecture for security functions with minimal packet loss, *(ii)* conducting extensive measurements applying hardware-supported timestamping to precisely determine worst-case latencies, and *(iii)* introducing a model to predict the capacity of our low-latency system for overload prevention. Our proposed system architecture relies on well-known applications and libraries such as Linux, the Data Plane Development Kit (DPDK), and Snort. Besides the specific measurements for the Snort IPS, we investigate the performance of the underlying operating system (OS) and libraries in use, namely Linux and DPDK, which emphasizes that our results are not limited to Snort but are highly relevant to other low-latency packet processing applications.

The remainder of the paper is structured as follows: Section II demonstrates the need for a new system design of security functions. Background and related work are presented in Section III. In Section IV we describe our novel system architecture which is evaluated in Section V. In Section VI we present our model for overload prediction. Considerations about the limitations and the reproducibility of our system architecture are given in Sections VII and VIII. Finally, Section IX concludes the paper by summarizing the most relevant findings and proposes enhancements for future work.

## II. MOTIVATION

The ITU [1] defines the requirements for the URLLC service as follows: the one-way delay of 5G radio access networks (RAN) from source to destination must not exceed 1 ms and the delivery success rate must be above 99.999%. We demonstrate how security functions facing input/output (IO) events, interrupts, and CPU frequency changes behave concerning the challenging URLLC requirements.

The following example uses Snort as an inline intrusion prevention system, i.e. every packet has to pass through Snort which subsequently influences the delay of every packet. For this example, all filtering rules are removed turning Snort into a simple forwarder which is not influenced by any rule processing. Therefore, the observed behavior represents a best-case scenario providing a lower latency bound for Snort IPS execution. Snort runs in a Virtual Machine (VM) providing a realistic multi-tenant setup for 5G networks. The packet rate is set to 10 kpackets/s, a moderate system load without any packet drops. The measurement runs 30 s. As we are not interested in the latency spikes caused by the application start-up, we exclude the first second of measurements from Figure 1 and Table 1.

TABLE I: Latencies of a Snort Forwarder

|     | n-*th* percentiles | | | | |
| --- | --- | --- | --- | --- | --- |
|     | 50 | 99 | 99.9 | 99.99 | 99.999 |
| Fwd | 69 µs | 88 µs | 107 µs | 1.7 ms | 2.5 ms |

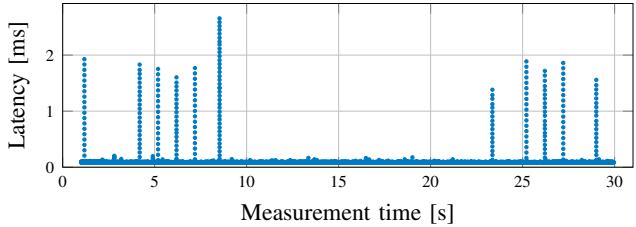

Fig. 1: Snort forwarder worst-case latencies

Table I shows the percentiles of latency we measured. Up to the 99.9th percentile, the observed latency budget is low enough to allow additional packet processing tasks while still meeting URLLC requirements. For higher percentiles, the latency budget is already exceeded by the basic forwarder. Figure 1 shows a scatter plot displaying the 5000 worst-case latencies measured over 30 s. We see that latencies exceeding the 1 ms latency budget are not only occurring at the beginning of a measurement due to cache warm-up or other ramp-up effects but in an irregular and unpredictable pattern throughout the entire measurement. The latency spike pattern did not change over time, therefore, we consider this being the steady-state behavior of our investigated application.

Thus, a different system design for security functions is needed to meet the strict requirements of URLLC services. In the following, we demonstrate techniques and frameworks creating a low-latency software stack to show that meeting URLLC requirements is possible while using the same hardware as for this motivating example.

## III. BACKGROUND AND RELATED WORK

After introducing the challenges of 5G, this section focuses on techniques impacting the delay and jitter caused by software packet processing systems.

*a) URLLC in Industry 4.0:* 5G has a strong focus on industrial Internet of Things (IoT) like Industry 4.0. These industrial networks will be highly heterogeneous in the sense of functionality, performance, and security protection. Network slicing is considered as a mechanism to handle the diverse set of requirements to the network and it is a challenge to construct slices with strict end-to-end (e2e) latency and reliability guarantees: industrial applications like control and alarm systems may require an e2e-latency in the range of 0.5 to 5 ms [3]. To address the problem, wireless URLLC is introduced in 5G and requires a new system design for the radio access network, called new radio (NR). The 3rd Generation Partnership Project (3GPP) standardization organization studied the URLCC requirements for NR [4]. But URLLC is not just about new radio. Besides the importance of latency and jitter, URLLC use cases like factory automation, remote robotics, and smart city automation have stringent requirements on availability [5]. Thus, URLLC is reflected in new concepts for operation and maintenance (O&M) including intelligent monitoring of network failures and cybersecurity attacks, surveillance, and instant actions.

Today, solutions to perform automated security management in and across slices while guaranteeing performance and security requirements are missing. Our target is the definition of an architecture and mechanisms for security (monitoring) functions to guarantee the QoS during design, deployment, and modification of slices for URLLC use cases.

*b) Polling vs. interrupts:* One possible cause for interrupts in an OS is the occurrence of IO events, such as arriving packets, to be handled by the OS immediately. This may cause short-time disruptions for currently running processes due to interrupt handling. The ixgbe network driver and Linux employ moderation techniques to minimize the number of interrupts and therefore the influence on processing latency [6]. Both techniques were introduced as a compromise between throughput and latency optimization. For our low-latency design goal, neither technique is optimal, as the interrupts—although reduced in numbers—still lead to irregular variations in the processing delay which should be avoided. DPDK [7], a framework optimized for high-performance packet processing, prevents triggering interrupts for network IO entirely. It ships with its own userspace driver avoiding interrupts, instead packets must be polled actively. This leads to a more predictable performance with little variation for execution times. Execution times are stabilized further due to DPDK's preallocation of memory and a lack of costly context switches between userspace and kernelspace. However, active polling requires the CPU to wake up regularly increasing energy consumption.

*c) CPU features:* Numerous guides list CPU and OS features leading to unpredictable behavior for application performance on which the following recommendations are based on [8]–[10]. HyperThreading (HT) or simultaneous multithreading (SMT) is a feature of modern CPUs which allows addressing physical cores (p-cores) as multiple virtual cores (v-cores). Each p-core has its own physically separate functional units (FU) to execute processes. However, multiple v-cores are hosted on a p-core sharing FUs between them. Zhang et al. [11] demonstrate that sharing FUs between v-cores can impact application performance when executing processes on v-cores instead of the physically separate p-cores. Another feature of modern CPUs are sleep states which lower CPU clock frequency and power consumption. Switching the CPU from an energy-saving state to an operational state leads to wake-up latencies. Schöne et al. [12] measured wake-up latencies between 1 µs and 40 µs for Intel CPUs depending on the state transition and the processor architecture.

Despite having physically separate FUs, p-cores share a common last level cache (LLC). Therefore, processes running on separate p-cores can still impact each other competing on the LLC. Herdrich et al. [13] observed a performance penalty of 64% for a virtualized, DPDK-accelerated application when running in parallel with an application utilizing LLC heavily. The uncontended application performance can be restored for the DPDK application by dividing the LLC statically between CPU cores utilizing the cache allocation technology (CAT) [13] of modern Intel CPUs.

*d) OS features:* Besides interrupts caused by IO events, an OS uses interrupts for typical tasks such as scheduling or timers. Patches for the Linux kernel [14] were introduced to create a more predictable behaving kernel, e.g. by reducing the interrupt processing time. Major distributions, such as Debian, provide this, so-called PREEMPT_RT kernel as part of their package repository. In addition, the Linux kernel offers several command line arguments influencing latency behavior. Cores can be excluded from the regular Linux scheduler via `isolcpu`. Isolated CPU cores should be set to `rcu_nocb` lowering the number of interrupts for the specified cores.

*e) Low-latency VM IO:* Transferring packets into/out of a VM leads to significant performance penalties compared to bare-metal systems. Emmerich et al. [2] compared packet forwarding in bare-metal and VM scenarios, demonstrating that VMs can introduce high tail latencies of 350 µs and above. They also demonstrated that DPDK can help improving forwarding latencies but must be used on the host system and the VM. Furthermore, modern network interface cards (NICs) supporting single root IO virtualization (SR-IOV) can be split into several independent virtual functions, which can be used as independent NICs and can be bound to VMs exclusively. In this case, virtual switching is done on the NIC itself, minimizing the software stack involved in packet processing. In an investigation by Lettieri et al. [15] SR-IOV, among other techniques for high-speed VM-based network functions, is one of the fastest techniques with the lowest CPU utilization. Therefore, latency performance of SR-IOV is superior to software switches, e.g. Xu and Davda [16] measured an almost 10-fold increase of worst-case latencies for a software switch. Xiang et al. [17] create and evaluate an architecture for low latency network functions. Their architecture provides sub-millisecond latencies, but they do not investigate the worst-case behavior. Zilberman et al. [18] give an in-depth latency analysis of various applications and switching devices. They stress the need for tail-latency analysis to comprehensively analyze application performance.

The topic of VM-based network functions has been extensively researched in literature [15]–[17]. However, given our motivating example in Section II and the importance of the URLLC service, we argue, similar to Zilberman et al. [18], that the crucial worst-case behavior needs close attention. In this work, we aim to create the lowest latency system achievable utilizing available applications running on off-the-shelf hardware.

There are also embedded systems such as jailhouse [19] or PikeOS [20] being able to partition the available hardware providing real-time guarantees for user processes or VMs. However, they are either not compatible with standard Linux interfaces such as libvirt or replace the host OS entirely.

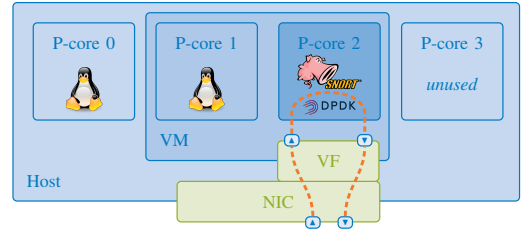

Fig. 2: Architecture overview

Therefore, the tool support for these specialized hypervisors is worse compared to the more widespread solutions such as Xen or KVM utilizing the libvirt software stack. Thus, we do not consider these specialized solutions for this work but rely on well-established software tools and hardware.

## IV. ARCHITECTURE DESIGN

Figure 2 shows the design of our low-latency VM running on a CPU with four physical cores (p-cores). To avoid any influence of v-cores, SMT is disabled for this system. A PREEMPT_RT kernel runs on the host and the VM to minimize interrupt latencies for the virtualized packet processing application. We use the core isolation feature of Linux to dedicate cores to specific processes aiming to minimize the QoS impact between cores and applications running on them. The OS of the *host* is running on p-core 0 exclusively, with kernel arguments isolating p-cores 1 and 2 for exclusive VM usage. On the *VM* itself, the OS is running on p-core 1 exclusively, with its kernel arguments set to p-core 2 being isolated. P-core 2, isolated from both host and VM, runs the application relying on DPDK and Snort. The core isolation feature complements DPDK's design philosophy of statically pinning packet processing tasks to cores. Utilizing SR-IOV, the NIC is split into virtual functions (VF). One VF is passed through to the VM being attached to p-core 2. The critical network path and its associated CPU resources are isolated from OS operation to provide a stable service for latency-critical processes. We disable the energy-saving states in the bios or set them to the most reactive state to avoid any delays caused when waking up the CPU. Additionally, we use CAT to statically assign LLC to cores.

## V. EVALUATION

The following measurement series characterizes the latency behavior of the proposed architecture.

### A. Setup

Figure 3 shows the setup used for testing based on three machines. The Device under Test (DuT) runs Snort forwarding traffic between its physical interfaces, the other two machines run the packet generator MoonGen [21]. The load generator (LoadGen) acts as traffic source/sink generating/receiving the test traffic, the third machine (timestamper) is monitoring the entire traffic received/sent by the DuT. The timestamper monitors the traffic between DuT and LoadGen via passive optical Terminal Access Points (TAP), timestamping every packet in hardware with a 12.5 ns resolution [22]. Being passive, the optical TAPs do not introduce variation to the timestamping process. Timestamping every single packet only works for receiving ports. Therefore, we timestamp on a separate host instead of the LoadGen itself.

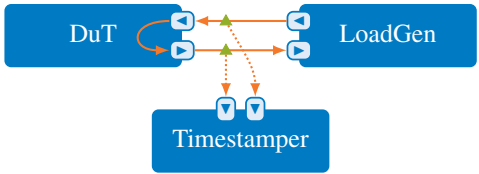

Fig. 3: Setup with Snort as a Device under Test (DuT), MoonGen as a LoadGen, and a Timestamper

The three servers are equipped with Supermicro mainboards (X10SDV-TP8F) featuring an Intel Xeon D-1518 CPU (4 cores, 2.2 GHz) and on-board Intel X552 NICs (dual-port SFP+, 10G Ethernet). On the DuT we use Debian buster (kernel v4.19) as OS, KVM as hypervisor, and the current beta of Snort (v3.0.0) [23] together with a DPDK-enabled data acquisition plugin (daq, v2.2.2) [24]. Section VIII lists the repositories, commit ids of the investigated applications, configuration data, and used measurement tools. The VM configuration is shown in Figure 2. Via Intel CAT [13], we pin 4 MiB of the LLC to the core running the packet processing application, the remaining 2 MiB are shared among the other cores.

We opt for UDP-only test traffic to prevent TCP congestion control from impacting the measured latency. The UDP destination port is set to 53 to trigger Snort's rules for DNS processing. The payload of our generated traffic does not contain DNS information but a counter to efficiently track packet loss and forwarding latency. We use constant bit rate traffic (CBR) for testing and dedicate Section V-D to measure the impact of bursty traffic.

### B. Measurements

The following measurements investigate the performance of our proposed architecture regarding the URLLC requirements. Therefore, we aim for a packet delivery rate above 99.999 % and a latency below 1 ms. We do not replicate the entire e2e communication path of 5G but only a security function located in the 5G backend network, therefore we aim for a lower latency goal. Faced with a similar problem Xiang et al. [17] calculated a latency goal of 350 µs for their network function chain. In this paper, we apply the same latency goal to our measurements quantifying the performance of Snort. We try to isolate the influence of the IO framework (DPDK), Snort overhead, and rule processing through separate measurements. Therefore, we test three related packet forwarding applications: *(i) DPDK-l2fwd*, being the most simple forwarder in our comparison, representing the minimum latency of IO without any processing happening; *(ii) Snort-fwd*, forwarding packet with Snort on top of DPDK, which quantifies the overhead caused by Snort without any traffic filtering happening; and *(iii) Snort-filter*, applying the Snort 3 community ruleset [25] to the forwarded traffic. The filter scenario does not drop any packet, because we are only interested in the overhead caused by rule application.

We measure between 10 and 120 kpackets/s incremented in 10 kpackets/s steps. Due to space limitations, we only show three selected rates for every scenario in Table II. For each scenario, we list the minimal rate of 10 kpackets/s, the last rate before overloading the DuT, and the first rate when the DuT was overloaded. The actual packet rates depend on the individual scenario. Being able to process millions of packets per second without overloading, we could not overload the DPDK forwarder within the selected packet rates [26]. Therefore, we present 10, 60, and 120 kpackets/s representing low, medium, and maximum load in this case.

*1) Hardware:* Initially, we test the forwarding applications in a non-virtualized setup to measure the performance baseline (cf. Table II, mode: HW).

*a) DPDK-l2fwd:* We measure the behavior of the DPDK forwarder for packet rates of 10, 60, and 120 kpackets/s. The median forwarding latency is 3.1 µs and increases slightly to a maximum of 3.4 µs for the 99th percentile indicating a stable latency behavior. Only the rare tail latencies, i.e. ≥ 99.9th percentile increase to a maximum value of 16.0 µs. The overall latency values do not differ significantly between measurements. We did not observe any packet loss for the three tested rates.

*b) Snort-fwd:* Running Snort on top of DPDK increases latency significantly. The median for rates of 10 and 80 kpackets/s is almost the same with 14.5 and 14.4 µs respectively. This new median is almost as high as the worst-case latency for the DPDK forwarder. Tail latencies increase further and seem to depend on the packet rate, i.e. tail latencies increase for higher packet rates. At a rate of 80 kpackets/s, packet drops can occur. A closer analysis showed that a consecutive sequence of packets was lost only at the beginning of the measurement, despite the previous warm-up phase. As packet loss did not occur later we do not consider this configuration as an overload scenario. We consider the rate of 90 kpackets/s as an overload scenario which is characterized by the noticeable packet loss (3.3%) and the over thousandfold latency increase compared to the median latency of the previous measurements. The latency increase in the overloaded scenario is the result of packets not being processed fast enough leading to buffers filling up. Therefore, the worst-case latency remains at this high level for all observed percentiles.

*c) Snort-filter:* For this measurement, the Snort forwarder applies the community ruleset. Rule application introduces additional costs resulting in a latency offset of roughly 3 µs compared to the previous measurement at 10 and 60 kpackets/s. Only the worst-case latencies differ noticeably for the latter. The overload scenario already occurrs at a lower rate of 70 kpackets/s due to the higher processing complexity indicated by the high tail latencies. Loss rates and median would still be tolerable, however, the tail latencies show an increase by a factor of over 1000 compared to the median.

TABLE II: Latencies of different software systems

| | Mode | Rate [kpackets/s] | Loss [%] | n-*th* percentiles | | | | | |
|---|---|---|---|---|---|---|---|---|---|
| | | | | 50 | 99 | 99.9 | 99.99 | 99.999 | Max. |
| DPDK-l2fwd | HW | 10 | - | 3.1 µs | 3.4 µs | 7.7 µs | 12.2 µs | 13.4 µs | 13.6 µs |
| | HW | 60 | - | 3.1 µs | 3.3 µs | 8.3 µs | 13.5 µs | 14.4 µs | 16.0 µs |
| | HW | 120 | - | 3.1 µs | 3.3 µs | 8.1 µs | 13.2 µs | 14.3 µs | 14.6 µs |
| | VM | 10 | - | 3.3 µs | 4.0 µs | 14.7 µs | 17.6 µs | 19.0 µs | 19.2 µs |
| | VM | 60 | - | 3.3 µs | 4.0 µs | 15.4 µs | 18.9 µs | 19.9 µs | 21.5 µs |
| | VM | 120 | - | 3.3 µs | 3.9 µs | 16.6 µs | 20.2 µs | 21.3 µs | 22.9 µs |
| Snort-fwd | HW | 10 | - | 14.5 µs | 24.7 µs | 29.7 µs | 32.4 µs | 33.1 µs | 33.1 µs |
| | HW | 80 | 0.1 | 14.4 µs | 29.9 µs | 43.7 µs | 46.2 µs | 47.7 µs | 50.6 µs |
| | HW | 90 | 3.3 | 30 609.5 µs | 30 834.8 µs | 30 882.7 µs | 30 915.3 µs | 30 936.1 µs | 30 959.1 µs |
| | VM | 10 | - | 15.9 µs | 37.6 µs | 58.6 µs | 66.8 µs | 68.0 µs | 68.6 µs |
| | VM | 80 | 0.1 | 18.8 µs | 73.9 µs | 98.8 µs | 115.6 µs | 117.7 µs | 121.9 µs |
| | VM | 90 | 7.1 | 2469.6 µs | 2657.9 µs | 2679.8 µs | 2692.2 µs | 2700.6 µs | 2708.3 µs |
| Snort-filter | HW | 10 | - | 17.4 µs | 28.2 µs | 33.1 µs | 35.8 µs | 36.4 µs | 36.6 µs |
| | HW | 60 | 0.0 | 17.1 µs | 29.0 µs | 34.1 µs | 36.1 µs | 50.4 µs | 51.5 µs |
| | HW | 70 | 0.0 | 79.0 µs | 24 897.2 µs | 27 521.2 µs | 27 847.0 µs | 27 947.1 µs | 27 992.9 µs |
| | VM | 10 | - | 18.4 µs | 40.9 µs | 63.1 µs | 71.8 µs | 73.4 µs | 73.7 µs |
| | VM | 60 | 0.1 | 17.5 µs | 62.2 µs | 92.9 µs | 101.1 µs | 114.7 µs | 115.7 µs |
| | VM | 70 | 3.0 | 3036.9 µs | 3270.2 µs | 3294.4 µs | 3313.1 µs | 3326.8 µs | 3342.5 µs |

Note that when comparing the load scenarios before overload this forwarder processes packets with a lower latency than its respective counterpart for the Snort forwarder. We attribute this to the relative load which is higher for the Snort-fwd, i.e. it is more overloaded at 90 kpackets/s than the Snort-filter at a rate of 70 kpackets/s.

*2) Virtualization:* Processing packets in virtualized environments can have a significant impact on latency. To measure the impact, we repeated the previous measurements in a virtualized environment (cf. Table II (Mode: VM), Figure 2).

*a) DPDK-l2fwd:* In the virtualized environment, latency increases compared to the non-virtualized measurements. The median latency increases by 6%, but the tail latencies can increase by almost 60%. Table II shows that up to the 99th percentile packet rate has little influence on latency. For higher percentiles, a trend towards higher latencies seems to manifest.

*b) Snort-fwd:* Compared to its hardware counterpart latency increases by 30% for the median and up to more than 100% for the tail latencies. We observe the same initial packet loss in the non-overloaded case. Packet loss in the overloaded case is higher, as packet processing is more expensive in the VM for the same packet rate. In the overloaded case latencies are over 10 times lower compared to the hardware measurements still violating the 1 ms goal. Measurements show that enabling SR-IOV leads to the decrease for worst-case latencies due to smaller buffers, an observation confirmed by Bauer et al. [27].

*c) Snort-filter:* Comparing Snort-filter in virtualized and non-virtualized environments shows that the median latency increase is ≤ 1 µs. The values for the tail latencies increase by a factor of two or more for the virtualized environment.

Looking at Table II we can conclude that URLLC-compliant latency is only violated if the DuT is overloaded. Overload latencies rise by a factor of 1000 for the HW scenario and by a factor of 100 for the VM scenario. Without overloading the system the latencies are below URLLC requirements even for the most challenging scenario. When considering the worst-case scenario, Snort-filter (HW), a latency of 50.4 µs at the URLLC-required 99.999th percentile was measured. The overall observed worst-case latency for the VM scenario for the 99.999th percentile is 117.7 µs. Despite the latency difference of over 100% between HW and VM, both worst-case scenarios—HW and VM—do not violate our latency goal of 350 µs. In fact, the remaining latency budget allows for even more complex packet processing tasks.

## C. Tail latencies

Previously, we have shown that the measured tail latencies in the non-overloaded scenario do not impair URLLC latency goals. In this section, we want to investigate the effects causing the tail latencies to exclude potentially harmful consequences such as latency spikes or even short-term overload. Increased tail latencies are already present in the DPDK-l2fwd scenario indicating that their causes are already part of the basic packet processing steps. We investigate the differences between the bare-metal deployment and the virtual environment.

*a) HW:* We analyze the tail latencies in the non-overloaded scenarios. Figure 4a shows a scatter plot of the 5000 highest latency events measured over 30 s. The figure shows a horizontal line at approx. 3.4 µs, the area where the majority of latency events happen matching the 99th percentile given in Table II, Line 1. Above this horizontal line, a linear regular pattern over the 30-second measurement period is visible. We assume that the latency events above the horizontal line of 3.4 µs are a result of packets being delayed due to interrupt processing in the OS.

A closer investigation identified this pattern above the horizontal line as an interplay of two clocked processes—

OS interrupt generation on the DuT and generated CBR traffic pattern on the LoadGen. The pattern is created by an effect known as aliasing. Here we use the generated traffic as sampling process trying to detect interrupts. As the interrupts are too short ($\leq 13.6\,\mu s$) to be correctly detected at the generated traffic rate ($100\,\mu s$ inter packet gap) we undersample leading to the observed pattern. The OS interrupt counters (`/proc/interrupts`) revealed local timer interrupts (loc) and IRQ work interrupts (iwi) to be the only interrupts triggered on the packet processing core of the DuT during operation. We measured, using the time stamp counter (TSC) of our CPU, constant execution times of $8.2\,\mu s$ for the iwi and $5.5\,\mu s$ for the loc. The two different execution times are visible in Figure 4a as longer and shorter lines. Their maximum values of 10.9 and $13.6\,\mu s$ differ because additional tasks such as packet IO and context switches are included. Packets were generated at a rate of $10\,kHz$, and we measured interrupts being generated at a rate of $250\,Hz$. Locs and iwis happen in a regular pattern, an iwi is triggered after every second loc.

To verify if the interrupts are the cause of the observed pattern, we create a script simulating the described process using the measured frequencies and processing times. Figure 4a shows similar patterns for the simulation confirming our assumptions. Measurement and simulation are highly sensitive to the measured maximum values, the traffic rates, and the interrupt rate. Even little parameter changes, e.g. restarting the load generator, can lead to changes in the generated traffic rate and therefore lead to different patterns. The same happens if the traffic rate is increased or lowered. This means that repeating the same measurements may lead to patterns with different shape and orientation. However, a regular pattern can be observed as long as the interrupt process is undersampled.

*b) VM:* Figure 4b shows the 5000 highest latency events measured for the DPDK-l2fwd (VM) scenario. The entire graph is shifted, the horizontal line is shifted to approx. $4.4\,\mu s$, the long interrupt latency is approx. $19\,\mu s$, the shorter approx. $16\,\mu s$, indicating the higher overhead when running in a VM. The number of events above the horizontal line roughly doubled. This can be explained that now two OSes (VM host and VM) trigger interrupts. We observed the same interrupts for the VM host as in our HW measurement, for the VM OS we only observed loc interrupts triggered at a rate of $250\,Hz$.

Despite our efforts to lower the number of interrupts by applying DPDK, there still remain a number of interrupts triggered by the OS itself causing latency spikes. Due to their scarcity and limited duration we do not consider them being harmful to our pursuit of building a latency-optimized system, when considering the URLLC latency goals.

### D. Influence of batch sizes

All previously described measurements use CBR traffic. For CBR the pauses in between packets can be used for packet processing without delaying subsequent packets and leading to optimal latency results. However, real traffic may arrive in bursts of packets without pauses between them. There, packet processing may delay subsequent packets. The

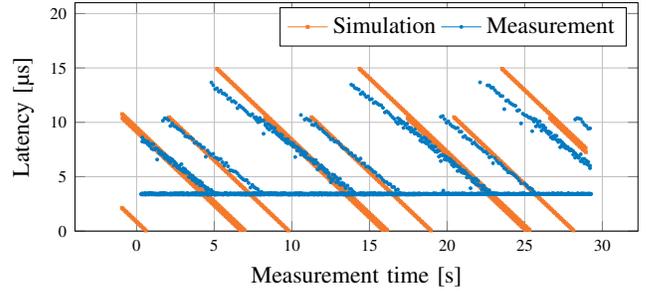

(a) HW (cf. Table II, Line 1)

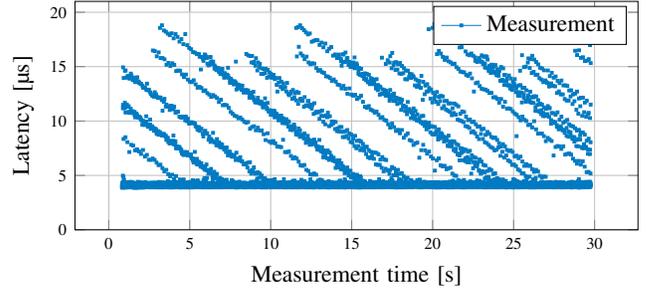

(b) VM (cf. Table II, Line 4)

Fig. 4: 5000 worst-case latency events measured for DPDK-l2fwd at 10 kpackets/s

following measurements show the impact of bursty traffic on the latency.

We define a block of packets arriving back-to-back on the wire as a burst and a block of packets being accepted or processed on a device as batch. Batched packet processing leads to a higher throughput for packet processing frameworks like DPDK [26]. The DPDK-enabled Snort accepts batches of up to 32 packets, processes them, and then releases the batch of packets only after all batched packets have been processed. Figure 5a shows the results of this processing strategy for different batch sizes. All graphs show areas of very steep increases indicating a large number of packets sharing the same latency, i.e. a batch of packets is sent out. Starting with batch size 4 flat areas become visible indicating that no packets were observed with this latency, i.e. the batch is processed without any packet sent out. The flat areas are followed by steep increases where the batches are sent out and the flat areas grow with increasing batch sizes as batch processing times increase. For a 64-packet burst, a two-step pattern is observed as two batches of 32 packets are processed in sequence. The plots show few packets with lower latency for every burst size. This happens if only a few packets of a burst are put into a batch, processed, and sent out before the remaining packets of the burst are processed.

As already processed packets are delayed until the batch is fully processed, the median delay is raised significantly. For low-latency optimized systems, smaller batch sizes can be beneficial. Therefore, we change the batch size from 32 to the minimal DPDK-supported batch size of 4. The results can be seen in Figure 5b, where the CDFs display a linear trend for growing burst sizes. This results in a significantly lower median for burst sizes of 16 and above with little influence on the maximum observed latency.

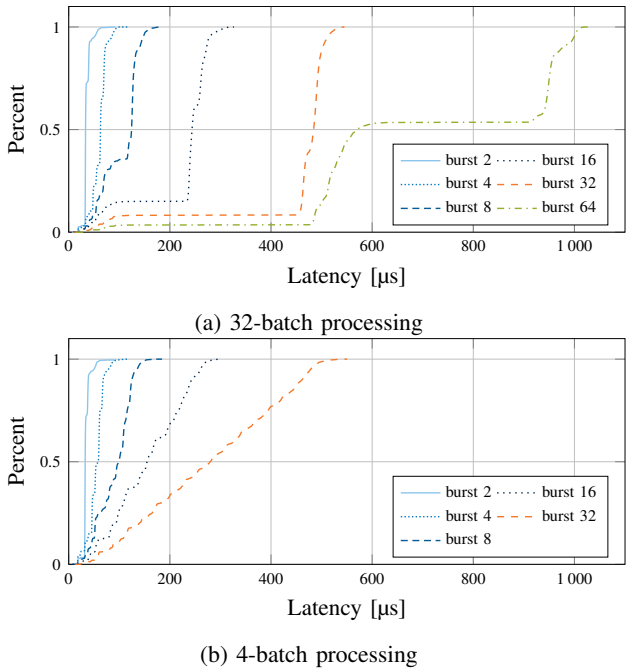

(a) 32-batch processing

(b) 4-batch processing

Fig. 5: Latency when forwarding using Snort-filter (VM) at 10 kpackets/s for different burst sizes

We have shown that the blocking behavior of this batch-processing strategy may increase latency unnecessarily. For low-latency systems, small batch sizes or even no-batch processing decrease latency. However, large bursts may cause latency violations due to short-time overload scenarios. In our case, burst sizes of 32 and 64 lead to latencies not meeting the URLLC criteria any longer for the chosen scenario.

*E. Energy consumption*

Our proposed low-latency configuration requires deactivating energy-saving mechanisms. Therefore, we compare the system configuration used for testing with a configuration with default bios settings and kernel arguments for energy saving enabled. For the power measurement, we use the metered power outlet Gude Expert Power Control 8226-1. We measure the power consumption of the entire server.

Table III lists the measured power values. We observe no differences in power consumption between the different applications (DPDK, Snort-fwd, Snort-filter). We measure the server while idling, while the application is in an available state, and while the application is actively processing packets. With power saving enabled there is a 14-watt difference between idle and transmitting state and a 3-watt difference between running and transmitting. The latter, rather low difference is a consequence of DPDK's design, relying on active polling, therefore, keeping the system (re-)active even without packet transfer. This intentional design decision of DPDK makes it a well-suited framework for high-performance scenarios where energy consumption is always high, however, DPDK is a poor choice for scenarios with low load because of the high energy consumption.

TABLE III: Power consumption

| Power saving | Idle | Available | Processing |
| --- | --- | --- | --- |
| enabled | 31 W | 42 W | 45 W |
| disabled | 46 W | 47 W | 47 W |

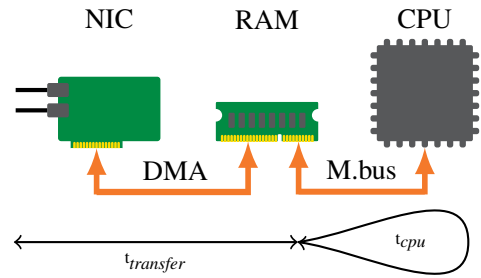

Fig. 6: Sources of delay on modern architectures

Disabling power saving increases the previous maximum power consumption by 1 W for the idle state and by 2 W for the other states. Comparing power saving enabled and disabled shows that low-latency configuration does not come for free. In scenarios with long idling periods, power consumption and therefore costs raise up to 48%. For other load scenarios the increase is lower (12% and 4%), i.e. if system load is already high for the traditional systems the additional costs for the low-latency configuration are significantly lower. The CPU used in our test system has a thermal design power (TDP) of 35 W, running more powerful CPUs with energy saving disabled may introduce even higher differences between idle and running states and therefore higher costs.

## VI. MODEL

Our measurements have shown that the packet processing system must not be overloaded to adhere to URLLC requirements. Therefore, we use our measurements to deduce a model calculating the maximum packet rate our system can handle without overloading.

Figure 6 shows the path of a packet through the different system components and their according time consumptions t. The time a packet travels through the system ($t_{transfer}$) includes delays caused by propagation, serialization, and the transfer from NIC to RAM. $t_{CPU}$ denotes the time the CPU processes the packet. As packets are received and sent, the path of a packet involves $t_{transfer}$ twice assuming symmetrical receiving and sending delays. This leads to Equation 1 for calculating the e2e delay of a single packet.

$$t_{e2e} = t_{cpu} + 2 t_{transfer} \quad (1)$$

In previous work [26] we demonstrated that the CPU is the main bottleneck in software packet processing. Especially considering the low packet rates ($\leq$ 120 kpacket/s) neither the

TABLE IV: Calculated CPU times and maximum rate

|  |  | DPDK-l2fwd $2t_{transfer}$ [µs] | Med. latency $t_{e2e}$ [µs] | CPUtime $t_{CPU}$ [µs] | Max. Rate $R_{max}$ [kpkts/s] |
|---|---|---|---|---|---|
| S.-fwd | HW | 3.1 | 14.5 | 11.4 | 87.4 |
|  | VM | 3.3 | 15.9 | 12.6 | 78.7 |
| S.-filter | HW | 3.1 | 17.4 | 14.3 | 69.7 |
|  | VM | 3.3 | 18.4 | 15.1 | 65.6 |

TABLE V: Trigger rates (r) & delays (d) of interrupts

|  | $loc_{host}$ | | $iwi_{host}$ | | $loc_{VM}$ | | $d_\Sigma$ per s |
|---|---|---|---|---|---|---|---|
|  | r [Hz] | d [µs] | r [Hz] | d [µs] | r [Hz] | d [µs] | [µs] |
| HW | 166.7 | 10.9 | 83.3 | 13.6 | - | - | 2949.9 |
| VM | 166.7 | 17.5 | 83.3 | 19.2 | 250 | 17.5 | 8891.6 |

Ethernet bandwidth, the NIC, nor the involved system busses are overloaded. Subsequently, it is crucial to determine the required calculation time on the CPU $t_{CPU}$ for calculating the maximum packet rate. Table II lists the measured e2e delays of the packets ($t_{e2e}$) not $t_{CPU}$. However, the DPDK-l2fwd scenario—representing the most basic forwarder possible without any processing for the packet—involves only a minimal amount of $t_{CPU}$. Therefore, we can use the median value of the DPDK-l2fwd measurement as a approximation of $2t_{transfer}$. Using that information, we can calculate an approximation of $t_{CPU}$ for Snort-fwd and Snort-filter, by deducting the median measured in the DPDK-l2fwd scenario from their respective e2e delays. The results of the approximated $t_{CPU}$ are given in Table IV.

Section V-C shows that a CPU core also performs interrupts. Depending on the scenario, the interrupt rates, and the costs of the individual interrupts Table V lists CPU time spent on interrupts per second $d_\Sigma$. Knowing the amount of CPU time spent on packet processing per packet $t_{CPU}$ and the CPU time spent on interrupt processing per second $d_\Sigma$ the Equation 2 can be deduced. Maximum packet rates calculated according to this equation are listed in Table IV.

$$R_{max} = \frac{1\,s - d_\Sigma}{t_{CPU}} \quad (2)$$

Comparing the calculated maximum rates in Table IV with the actual maximum rates measured in Table II shows that Equation 2 can predict the overload correctly for three out of four scenarios. For the Snort-fwd (VM) scenario the maximum rate is underestimated with the overload not happening at the predicted rate 78.7 kpackets/s but beyond 80 kpackets/s. We conclude that the prediction approximates a lower bound for the maximum packet rate. A conservative approximation is advisable in this scenario especially considering the devastating impact of overload on latency and QoS.

## VII. LIMITATIONS

Despite its benefits in terms of latency and jitter the proposed architecture has disadvantages. Statically assigning VMs to cores does not allow sharing a CPU core between several VMs, at least not the isolated cores dedicated to realtime applications. This increases hosting costs for such a VM. Migrating VMs or scaling the VM setup is not possible as SR-IOV does not allow VM migration due to the non-trivial replication of the NIC's hardware state [16]. Disabling energy-saving mechanisms increases energy costs for the server (cf. Section V-E), air conditioning, and increases the thermal load on the hardware which in turn may require earlier replacement additionally raising costs.

## VIII. REPRODUCIBILITY

As part of our ongoing effort towards reproducible network research, we release the pcap traces and plotting tools used in the measurements of this paper. Further, we release our measurement tools and source code of the investigated software including a detailed description for others to replicate our measurements on https://gallenmu.github.io/low-latency. Figures 1, 4 and 5 link to a detailed description of the used source code, experiment scripts, generated data, and plotting tools. The website lists additional measurements which were cut from the paper due to space limitations (cf. Table II).

## IX. CONCLUSION

Our paper shows that—in contrast to non-optimized systems—a carefully tuned system architecture can meet the demanding latency and reliability requirements of future 5G URLLC services. Hardware-timestamped latency measurements of the entire network traffic, allow for a detailed analysis of worst-case latencies, bursty traffic, and system load. We measured a virtualized system running a real-world intrusion prevention system causing a worst-case latency of 116 µs on a steady-state system, leaving enough room for subsequent packet processing tasks. Further, we show bursty traffic causing short-time overloads violating the latency requirements and introduce a strategy to reduce its impact. By publicly releasing our experiment scripts and data we provide the foundation for others to reproduce all measurements described in this paper.

We introduce a model to predict system overload to avoid the destructive effect of overload on latency. The benefits of this model are its simplicity requiring only the median forwarding latency for IO and the interrupt processing times.

Despite the increase in power consumption (48% for a low-load and 4% for a high-load scenario), we demonstrate that off-the-shelf hardware and available open-source software can achieve consistently low latency. Relying on off-the-shelf hardware and well-known tools simplifies the transition towards URLLC.

For future work, we want to investigate the impact of hosting different 5G service classes on the same system especially regarding potential QoS cross-talk and potential mitigation strategies.

## X. ACKNOWLEDGEMENTS

This work was supported by the Nokia university donation program and by the DFG-funded projects Moonshine & MoDaNet (grant no. CA 595/7-1 & CA 595/11-1).

## REFERENCES

[1] ITU. Report ITU-R M.2410-0 (11/2017) Minimum requirements related to technical performance for IMT-2020 radio interface(s). https://www.itu.int/dms_pub/itu-r/opb/rep/R-REP-M.2410-2017-PDF-E.pdf. Accessed: 2019-08-28.

[2] Paul Emmerich, Daniel Raumer, Sebastian Gallenmüller, Florian Wohlfart, and Georg Carle. Throughput and Latency of Virtual Switching with Open vSwitch: A Quantitative Analysis. July 2017.

[3] Philipp Schulz, Maximilian Matthe, Henrik Klessig, Meryem Simsek, Gerhard Fettweis, Junaid Ansari, Shehzad Ali Ashraf, Bjoern Almeroth, Jens Voigt, Ines Riedel, et al. Latency Critical IoT Applications in 5G: Perspective on the Design of Radio Interface and Network Architecture. *IEEE Communications Magazine*, 55(2):70–78, February 2017.

[4] Study on Scenarios and Requirements for Next Generation Access Technology. Technical report, 3GPP TR 38.913 V14.2.0, May 2017.

[5] Rapeepat Ratasuk. Ultra Reliable Low Latency Communication for 5G New Radio, October 2018. IEEE Workshop on 5G Technologies for Tactical and First Responder Networks.

[6] Paul Emmerich, Daniel Raumer, Alexander Beifuß, Lukas Erlacher, Florian Wohlfart, Torsten M. Runge, Sebastian Gallenmüller, and Georg Carle. Optimizing Latency and CPU Load in Packet Processing Systems. In *International Symposium on Performance Evaluation of Computer and Telecommunication Systems (SPECTS 2015)*, Chicago, IL, USA, July 2015.

[7] The Linux Foundation. Data Plane Development Kit (DPDK). https://dpdk.org. Accessed: 2019-08-28.

[8] Mark Beierl. Nfv-kvm-tuning. https://wiki.opnfv.org/pages/viewpage.action?pageId=2926179. Accessed: 2019-08-28.

[9] Joe Mario and Jeremy Eder. Low Latency Performance Tuning for Red Hat Enterprise Linux 7. http://people.redhat.com/jmario/docs/201501-perf-brief-low-latency-tuning-rhel7-v2.0.pdf. Accessed: 2019-08-28.

[10] AMD. Performance Tuning Guidelines for Low Latency Response on AMD EPYC-Based Servers. http://developer.amd.com/wordpress/media/2013/12/PerformanceTuningGuidelinesforLowLatencyResponse.pdf. Accessed: 2019-08-28.

[11] Yunqi Zhang, Michael A Laurenzano, Jason Mars, and Lingjia Tang. SMiTe: Precise QoS Prediction on Real-System SMT Processors to Improve Utilization in Warehouse Scale Computers. In *2014 47th Annual IEEE/ACM International Symposium on Microarchitecture*, pages 406–418. IEEE, December 2014.

[12] Robert Schöne, Daniel Molka, and Michael Werner. Wake-up Latencies for Processor Idle States on Current x86 Processors. *Computer Science-Research and Development*, 30(2):219–227, July 2015.

[13] Andrew Herdrich, Edwin Verplanke, Priya Autee, Ramesh Illikkal, Chris Gianos, Ronak Singhal, and Ravi Iyer. Cache QoS: From concept to reality in the Intel® Xeon® processor E5-2600 v3 product family. In *2016 IEEE International Symposium on High Performance Computer Architecture (HPCA)*, pages 657–668. IEEE, March 2016.

[14] Paul McKenney. A realtime preemption overview. https://lwn.net/Articles/146861/. Accessed: 2019-08-28.

[15] Giuseppe Lettieri, Vincenzo Maffione, and Luigi Rizzo. A survey of fast packet I/O technologies for Network Function Virtualization. In *International Conference on High Performance Computing*, pages 579–590. Springer, June 2017.

[16] Xin Xu and Bhavesh Davda. SRVM: Hypervisor Support for Live Migration with Passthrough SR-IOV Network Devices. In *ACM SIGPLAN Notices*, volume 51, pages 65–77. ACM, 2016.

[17] Zuo Xiang, Frank Gabriel, Elena Urbano, Giang T Nguyen, Martin Reisslein, and Frank HP Fitzek. Reducing Latency in Virtual Machines: Enabling Tactile Internet for Human-Machine Co-Working. *IEEE Journal on Selected Areas in Communications*, 37(5):1098–1116, 2019.

[18] Noa Zilberman, Matthew Grosvenor, Diana Andreea Popescu, Neelakandan Manihatty-Bojan, Gianni Antichi, Marcin Wójcik, and Andrew W Moore. Where Has My Time Gone? In *International Conference on Passive and Active Network Measurement*, pages 201–214. Springer, February 2017.

[19] Ralf Ramsauer, Jan Kiszka, Daniel Lohmann, and Wolfgang Mauerer. Look Mum, no VM Exits! (Almost). *CoRR*, abs/1705.06932, May 2017.

[20] Robert Kaiser and Stephan Wagner. Evolution of the PikeOS Microkernel. In *First International Workshop on Microkernels for Embedded Systems*, volume 50, January 2007.

[21] Paul Emmerich, Sebastian Gallenmüller, Daniel Raumer, Florian Wohlfart, and Georg Carle. MoonGen: A Scriptable High-Speed Packet Generator. In *Internet Measurement Conference (IMC) 2015, IRTF Applied Networking Research Prize 2017*, Tokyo, Japan, October 2015.

[22] Intel. Ethernet Controller X550. https://www.intel.com/content/dam/www/public/us/en/documents/datasheets/ethernet-x550-datasheet.pdf?asset=12457, 11 2018. Accessed: 2019-08-28.

[23] Cisco Inc. Snort. https://github.com/snort3/snort3. Accessed: 2019-08-28.

[24] Napatech. Snort. https://github.com/napatech/daq_dpdk_multiqueue. Accessed: 2019-08-28.

[25] Talos et al. Snort3 community ruleset. https://www.snort.org/downloads/#rule-downloads. Accessed: 2019-08-28.

[26] Sebastian Gallenmüller, Paul Emmerich, Florian Wohlfart, Daniel Raumer, and Georg Carle. Comparison of Frameworks for High-Performance Packet IO. In *ACM/IEEE Symposium on Architectures for Networking and Communications Systems (ANCS 2015)*, Oakland, CA, USA, May 2015.

[27] Simon Bauer, Daniel Raumer, Paul Emmerich, and Georg Carle. Intra-node Resource Isolation for SFC with SR-IOV. In *IEEE 7th International Conference on Cloud Networking (CloudNet'18)*, Tokyo, Japan, October 2018.